\documentclass[aps,prl,twocolumn,groupedaddress,showpacs,preprintnumbers]{revtex4}

\usepackage{graphicx}

\begin{document}
\title{Beyond Boltzmann-Gibbs statistics: Maximum entropy hyperensembles out-of-equilibrium}

\author{Gavin E. Crooks }
\email{gecrooks@lbl.gov}
\homepage{http://bespoke.lbl.gov/}
\affiliation{Physical Bioscience Division, Lawrence Berkeley National Laboratory, Berkeley, California 94720}
\date{\today}
\preprint{LBNL-59723}

\begin{abstract}
What is the best description that we can construct of a thermodynamic system that is not in equilibrium, given only one, or a few, extra parameters over and above those needed for a description of the same system at equilibrium? Here, we argue the most appropriate additional parameter is the non-equilibrium entropy of the system, and that  we should not attempt to estimate the probability distribution of the system, but rather the metaprobability (or hyperensemble) that the system is described by a particular probability distribution. The result is an entropic distribution with two parameters, one a non-equilibrium temperature, and the other a measure of distance from equilibrium. This dispersion parameter smoothly interpolates between certainty of a canonical distribution at equilibrium and  great uncertainty as to the probability distribution as we move away from equilibrium. We deduce that, in general, large, rare fluctuations become far more common as we move away from equilibrium.    
\end{abstract}
\maketitle

\begin{figure}[t]
\includegraphics{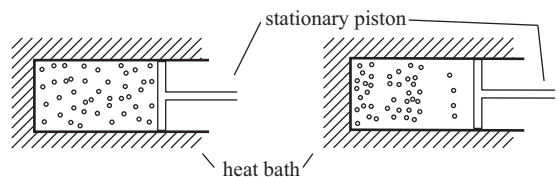} 
\caption{
Schematic realizations of a gas confined to a piston in and out of equilibrium.
}
\label{pistonfig}
\end{figure}

Consider a gas confined to a piston,  as illustrated in figure~\ref{pistonfig}. The realization on the left was sampled from thermal equilibrium with a fixed plunger. To describe the probability of every single possible configuration of the particles we only need to know the Hamiltonian of the system and the temperature of the environment~\cite{Gibbs1902}.  On the other hand, the system on the right has been sampled from a non-equilibrium ensemble. Although the Hamiltonian is the same, the plunger has recently been in violent motion, and this perturbation has driven the ensemble away from equilibrium. 
  To describe the configurational probability we now need to know the entire past history of  perturbations that the system  has undergone.  The dynamics and historical details matter.

This example illustrates the essentially difficultly we face when trying  to directly extend equilibrium statistical mechanics out of equilibrium. There is only one ensemble that can describe a given system in thermal equilibrium, but there are a multitude of ways that the same system can be out-of-equilibrium.  
The constraint that the equilibrium entropy is maximized is a very strong condition. 
However, let us take a step back, and reflect that statistical mechanics itself is designed to circumvent a similar difficulty.
In classical mechanics we typically assume that we know the exact microstate of the system. However, in statistical mechanics, we recognize that often such a detailed description is  neither possible nor desirable. 
A few bulk measurements  or parameters do not provide nearly enough information to fix the microstate. Instead we content ourselves with calculating the probability that the system occupies a particular microstate. To ask what the state of the system \emph{is}, rather than what it \emph{could be}, is to ask an unnecessarily difficult question. 

Out-of-equilibrium we essentially face the same problem, compounded. Clearly we cannot obtain enough information from a few measurements to determine the microscopic state of the system, but if the system is out of equilibrium then a few parameters or measurements are not sufficient (in general) to determine the ensemble either. Therefore, perhaps the correct approach is not to try to determine what the probability distribution of the system is, but instead attempt to determine what the probabilities could be.
In other words, instead of thinking about an ensemble of systems, we instead envisage an ensemble of ensembles, a `hyperensemble', where each member of the hyperensemble 
has the same instantaneous Hamiltonian, but is described by a different probability distribution.
We seek a generic  description of  the typical non-equilibrium ensemble given a few parameters or measurements that describe the average behavior of the hyperensemble.

This basic approach is borrowed from  Bayesian statistics, where it is not uncommon to estimate the probability of a probability density (a `metaprobability') when the available data is too sparse to reliable estimate the probability directly~\cite{Jaynes2003,DurbinEddy1998,GelmanCarlin2003}. Reference~\cite{DurbinEddy1998} contains a lucid description of this procedure in the context of amino acid sequence profiles.
 The hyper- prefix is also borrowed from Bayesian statistics, were it is usual to talk about hyperpriors (a prior distribution of a prior distribution) and associated 
 hyper-parameters.

With this insight, we can  move beyond the standard canonical ensemble by changing the question.
Instead of trying to find the probability distribution $\theta$ of the system directly, we instead estimate the metaprobability $P(\theta$),
the probability of the microstate probability distribution. 
We proceed analogously to the maximum entropy derivation 
of equilibrium statistical mechanics~\cite{Jaynes1957a, Gibbs1902}.
We will find the probability distribution of ensembles $P(\theta)$ that maximizes the entropy~$\mathcal H$ of the
hyperensemble,
\begin{equation}
	{\mathcal H}\Big(P(\theta) \Big) = - \int P(\theta) \log \frac{P(\theta)}{ m(\theta)} \;   d\theta \;,
\label{hyperentropy}
\end{equation}
while maintaining certain appropriate constraints. Here, $m(\theta)$ is a measure on the space of
probability distributions. It acts as a prior and ensures that this entropy is invariant under a change of variable. 

The trick to maximum entropy methods is finding the appropriate constraints, since with an arbitrary choice
of constraint  and prior practically any answer can be manufactured. To avoid this trap, we seek a minimal set of 
physically and mathematically reasonable parameters. Clearly, the hyperensemble must be normalized,
\begin{equation}
	1 = \int P(\theta) \; d\theta  \; .
\label{HN}
\end{equation}
And, by analogue with the canonical ensemble, we should constrain the mean energy of the ensemble of ensembles, 
\begin{equation}
\Big\langle\langle E \rangle\Big\rangle = \int  P(\theta)  \Big[\sum_i \theta_i E_i\Big]  \; d\theta
\label{ME}
\end{equation}

Thus far, we have incorporated the same information and constraints that lead to the canonical ensemble, namely the density of energy states, normalization and mean energy. To move beyond the canonical ensemble we require a measure of how far the system is from equilibrium. After all, the quintessential feature of non-equilibrium systems is that they are not in equilibrium. What is the most appropriate measure? If the system were in equilibrium, then the entropy would be maximized given the constraints. It follows that out-of-equilibrium the entropy of the ensemble is not maximized, and moreover, the entropy cannot be determined with any certainty from a measurement of the mean energy alone.  Therefore, the entropy itself can be used as an additional, physically relevant constraint.
\begin{equation}
\langle S \rangle = \int P(\theta) \;  \left[-\sum_i \theta_i \log \theta_i\right] d\theta \;  
\label{meanentropy}
\end{equation}

To summarize, we will maximize the entropy of the hyperensemble (Eq.~\ref{hyperentropy}) subject to normalization, the mean energy and the mean ensemble entropy (Eq.~\ref{HN}--\ref{meanentropy}).  The solution to this problem is found by introducing Lagrange multiplies $\{\lambda\}$  and then applying the calculus of variation  in the usual way:
\begin{equation}
P(\theta) = m(\theta) e^{-\lambda_0 -\lambda_1 \langle E \rangle - \lambda_2 S(\theta)} 
\end{equation}

Some manipulation will illuminate the significance of this expression. Let us rewrite with $\lambda_0 = \log \mathcal Z$, $\lambda_1 = \lambda \beta$ and $\lambda_2 = \lambda$.
\begin{equation}
P(\theta) = \frac{m(\theta)}{{\mathcal Z}(\beta,\lambda) }\exp\left({-\beta\lambda \sum_i \theta_i E_i + \lambda \sum_i \theta_i \log \theta_i }\right)
\end{equation}
In the absence of any compelling evidence to the contrary,
we will assume a uniform, uninformative prior over probabilities, $m(\theta) \propto$~constant. 
The parameter $\beta$ has units of entropy per unit of energy and is  effectively an inverse temperature.
Therefore, we can 
naturally introduce a canonical ensemble with the same effective temperature, 
\begin{equation}
\rho_i = \frac{1}{Q(\beta)}\exp(-\beta E_i) ,
\label{reference}
\end{equation}
and rewrite the maximum entropy hyperensemble as 
\begin{equation}
\left\| \qquad
P(\theta) = \frac{Q(\beta)}{{\mathcal Z}(\beta,\lambda) }\exp\left({-\lambda \sum_i \theta_i \log \frac{\theta_i}{\rho_i} }\right)
\qquad
\right.
\label{CHE}
\end{equation}
It is now evident that our hyperensemble has the functional form of the entropic distribution, a probability of probabilities that occasionally occurs in Bayesian statistics~\cite{Skilling1989,Skilling1990,Rodriguez1989, Caticha2001,Caticha2004}. 
This same functional form also appears as the asymptotic limit of the multinomial distribution with large sample sizes~\cite{CoverThomas1991},  in large deviation theory~\cite{CoverThomas1991, Ellis1999}, and as the natural conjugate prior of the Dirichlet distribution. 

\begin{figure}
\includegraphics{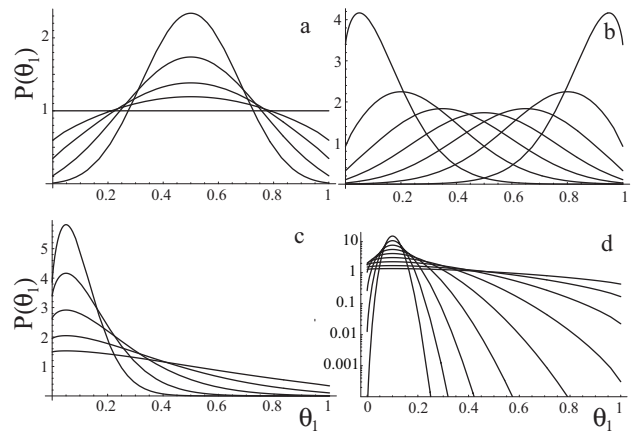} 
\caption{The entropic distribution (Eq.~\ref{CHE}) over 2 states. (a) Reference distribution $\rho=(0.5,0.5)$, $\lambda=0,1,2,4,8$ (Broad to peaked) (b) $\lambda=4$, $\rho_1 = 0.05, 0.20, 0.35, 0.5, 0.65, 0.80, 0.95$ (left to right). (c) $\rho=(0.1,0.95)$, $\lambda=0.5,1,2,4,8$. (d) Same $\rho$, log scale, $\lambda=0.5,1,2,4,8,16,32,64,128$. Note that the reference distribution controls the mode and that as the dispersion parameter $\lambda$ approaches $0$ the distributions become broader and the mean moves towards $\frac{1}{2}$. 
}
\label{entropic}
\end{figure}

\begin{figure}
\includegraphics{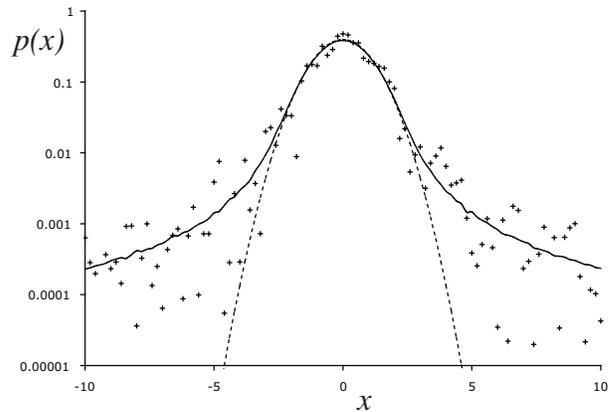} 
\caption{The entropic distribution (Eq.~\ref{CHE}) with a Gaussian reference distribution (zero mean, unit variance) and
dispersion $\lambda=100$. 
The dashed line is the reference $\rho$, the points are a single Monte-Carlo sample of $\theta$, and the
 solid line is the mean distribution $\langle\theta\rangle$. Note that the variation of $\theta$ away  from
 the reference is relatively large for intrinsically rare states, $\rho < 1/\lambda$.
}
\label{gaussian}
\end{figure}

The entropic distribution over a binary state space is illustrated in fig.~\ref{entropic}, and with a Gaussian reference
(e.g. a particle in a harmonic potential)
 in fig.~\ref{gaussian}.
We see that as $\lambda$ decreases the dispersion of the probability distributions increases, the mean distribution moves away from the canonical distribution, the average probability of rare states increases, and the probability of common states decreases to compensate. Moreover, in fig.~\ref{gaussian} we see that $\lambda$ controls a crossover in behavior; if $\rho > \lambda^{-1}$ then the uncertainty in $\theta$ and the bias away from equilibrium are relatively small, whereas for rare states, $\rho< \lambda^{-1}$, the perturbation are large. 
Therefore, the generic, predicted behavior is that rare events typically (but not necessarily) become far more common as the condition of thermal equilibrium is relaxed.

We can deduce some important properties of the hyperensemble by  noting that  
the function in the exponential of Eq.~\ref{CHE} is the relative entropy of $\theta$ 
to the reference canonical distribution,~$\rho$~\cite{CoverThomas1991}:
\begin{equation}
D(\theta \| \rho) = \sum_i \theta_i \log \frac{\theta_i}{\rho_i}
\label{relative}
\end{equation}
This is a natural measure of how distinguishable one distribution is from  another.
Since the relative entropy is zero if the distributions are identical, and 
positive if they are not, 
it immediately follows that the mode of the entropic distribution 
is located at the reference $\rho$. In other words, the single most 
probable distribution of the hyperensemble is a canonical distribution controlled 
by the effective temperature $\beta$, and the dispersion of the hyperensemble about 
that mode is controlled by the inverse scale parameter $\lambda$. If $\lambda$ 
is very large the hyperensemble collapses to a single point at the mode and we
recover the canonical ensemble of equilibrium statistical mechanics.  
It follows that
the reference temperature is numerically equal to the conventional 
temperature of the same system with the same mean energy at thermal equilibrium.
As $\lambda$ decreases
the dispersion increases and typical distributions differ significantly from the reference, 
until at $\lambda=0$ every distribution in equally likely.

Another way of  looking at the canonical hyperensemble is to note that the relative entropy of $\theta$ to a canonical reference $\rho$ can be interpreted as 
a generalized free energy difference~\cite{Qian2001}.
\begin{eqnarray}
	D(\theta \| \rho) &=& \beta F(\theta)  -\beta F(\rho)   ,\\
	\beta F(p) &=& -\sum_i p_i \log p_i + \beta \sum_i p_i E_i \nonumber
\end{eqnarray}
Since $\rho$ is canonical $F(\rho)= S/\beta -  \langle E \rangle$ is the Helmholtz free energy, whereas $F(\theta)$ can be interpreted as a
 generalized, non-canonical
free energy. Using these definitions, the canonical hyperensemble  can be written as  
\begin{equation}
P(\theta) \propto \exp \left\{   - \lambda \beta \big[F(\theta)  - F(\rho)\big]   \right \} \; .
\end{equation}
The physical picture is that near thermal equilibrium the ensemble that maximizes the free energy dominates the hyperensemble. As we move away from equilibrium the free energy is no longer necessarily maximized. Rather the probability of obtaining a particular ensemble out of equilibrium is determined by the generalized
free energy difference between that ensemble and the reference canonical ensemble.
This expression is pleasingly reminiscent of the thermodynamic fluctuation representation of standard statistical mechanics~\cite{Callen1985}, except we are now looking at  fluctuations in ensemble rather than state.

We can also derive the entropic hyperensemble by directly constraining the mean relative entropy $\langle D(\theta\|\rho)\rangle$.
From the viewpoint of  information theory, this is the average penalty for encoding states of the system assuming the they are drawn from the reference distribution $\rho$ rather than the true distributions~\cite{CoverThomas1991}. 
This measure is very similar to the Jensen-Shannon divergence
$\langle D(\theta\|\langle\theta\rangle)\rangle$~\cite{Lin1991}, except that
the reference distribution is the mode, rather than the mean of $\theta$. 

Currently, various modifications or extensions of Boltzmann-Gibbs statistics are being investigated,
including Tsallis statistics (Which modifies the entropic function)~\cite{Tsallis1988} and maximum entropy
production (Which modifies the constraints)~\cite{Dewar2003}.
Perhaps the most similar approach to the present work is
superstatistics~\cite{BeckCohen2003}, the central idea of which is  
that a system may be locally in equilibrium (either in time or space), but globally out-of-equilibrium. Therefore,
the system as a whole can be described by a \emph{mixture} of canonical ensembles, each with a different
local temperature. In contrast, the components of the maximum entropy hyperensemble are not required
to be canonical.
The essentially difficulty with superstatistics is that the distribution of effective temperatures is
unconstrained.
It is therefore  interesting to ask what distribution of local temperature would maximize the
hyperentropy given that the members of the hyperensemble are canonical? 
 Since the result will depend on the density of states, let us explore a simple, but important, special case,
 a collection of harmonic oscillators. The partition function is $Q(\beta)=\beta^{-c}$ and therefore
 the mean energy scales as $\langle E \rangle = c/\beta$, where the constant `$c$' is proportional to the size of the system.
An obvious choice for the prior is $m(T)\propto 1/T$~\cite{Jaynes2003}. Plugging these relations into Eq.~\ref{CHE} we find
\begin{equation}
P(T) \propto \left( \frac{T}{T^{\circ}}\right)^{c\lambda-1 } e^{-c \lambda T/T^{\circ} }  \;,
\end{equation}
 where $T$ is the effective local temperature and $T^{\circ}=1/\beta$ is the reference temperature. 
 Here, with the hyperensemble approach we predict that if the system is linear and locally in equilibrium, then
the temperature fluctuations follow a
gamma distribution~\cite{Touchette2002,Sattin2002,Touchette2005} with mean $T^{\circ}$ and 
standard deviation $T^{\circ}/\sqrt{c\lambda}$. If the temperature fluctuations are not gamma distributed, then either the system is not linear, not in local equilibrium, or
we have failed to incorporate some important, pertinent information about the system~\cite{Jaynes2003}.

It is worth noting that we would have obtained very different results if we had chosen different constraints.
As previously mentioned, this is the essential weakness of maximum entropy methods; we must 
rely on the plausibility of the constraints, rather than the rigor of the derivation. 
In particular, if we maximize the hyperentropy given the 
mean relative entropy of the reference $\rho$ to the ensemble $\theta$,  
$\langle D(\rho\|\theta)\rangle$, we obtain a Dirichlet distribution. This in turn leads to the prediction that 
the local temperature of a linear system follows an \emph{inverse} gamma distribution, which is 
known to be equivalent to the non-extensive thermodynamics of Tsallis~\cite{BeckCohen2003,Tsallis1988}.
This is an intriguing connection, but unfortunately $\langle D(\rho\|\theta)\rangle$ has no immediately 
obvious deep physical or information theoretic significance.

In this paper, I have argued that a natural way of moving beyond equilibrium Boltzmann-Gibbs statistics is to change the
question: Instead of trying to determine what the probability distribution of a system \emph{is}, we instead ask
what the probability distribution \emph{could be}. We seek an ensemble of ensembles that captures the generic properties of matter generically out-of-equilibrium.
The solution to this problem is found
by maximizing the entropy of the hyperensemble, given the mean energy and mean ensemble entropy.
This yields a physically plausible description of fluctuations away from equilibrium, a natural definition 
of temperature out-of-equilibrium, a natural measure of distance away from equilibrium,
and the intuitively plausible prediction that rare events typically become far more common as
a system moves away from thermal equilibrium.

\begin{acknowledgments} 
This work was supported by the Director, Office of Science, of the U.S. Department of Energy under Contract No. DE-AC02-05CH11231.\end{acknowledgments}


\end{document}